\documentclass[usegraphicx,usenatbib,a4paper]{mn2e}
\usepackage{aas_macros,graphicx,times,multirow}
\usepackage{amssymb}
\usepackage{graphicx,natbib}
\usepackage{lscape}
\usepackage{float}
\usepackage[T1]{fontenc}
\usepackage{aecompl}
\title[The IP IGR J14257-6117]
{IGR J14257-6117, a magnetic accreting white dwarf with a very strong X-ray orbital modulation}
\author[F. Bernardini et al.]                                                    
{F.~Bernardini,$^{1,2,3}$\thanks{E-mail:federico.bernardini@oa-roma.inaf.it} 
D.~de Martino,$^{3}$ 
K.~Mukai,$^{4,5}$
M.~Falanga,$^{6}$
\\
$^1$ INAF - Osservatorio Astronomico di Roma, via Frascati 33, I-00040 Monteporzio Catone, Roma, Italy \\
$^2$ New York University Abu Dhabi, Saadiyat Island, Abu Dhabi, 129188, United Arab Emirates\\
$^3$ INAF $-$ Osservatorio Astronomico di Capodimonte, Salita Moiariello 16, I-80131 Napoli, Italy\\
$^4$ CRESST and X-Ray Astrophysics Laboratory, NASA Goddard Space Flight Center, Greenbelt, MD 20771, USA\\
$^5$ Department of Physics, University of Maryland, Baltimore County, 1000 Hilltop Circle, Baltimore, MD 21250, USA\\
$^6$ International Space Science Institute (ISSI), Hallerstrasse 6, CH-3012 Bern, Switzerland\\}

\date{}
\pagerange{\pageref{firstpage}--\pageref{lastpage}}
\def\Swift{{\em Swift}}

\def\XMM{{\em XMM-Newton}}

\def\ergscm{$\rm erg\,cm^{-2}\,s^{-1}$}
\def\INT{{\em INTEGRAL}\,}

\def\IGRJ{IGR\,J14257-6117}

\begin{document}

\label{firstpage}

\maketitle

\begin{abstract}

\IGRJ\ is an unclassified source in the hard X-ray catalogues. 
Optical follow-ups suggest it could be a Cataclysmic Variable of the magnetic type. We present the first high S/N X-ray observation performed by \XMM\ at 0.3--10 keV, complemented with 10--80 keV coverage by \Swift/BAT, aimed at revealing the source nature. 
We detected for the first time a fast periodic variability at 509.5\,s and a longer periodic variability at 4.05\,h, ascribed to the white dwarf (WD) spin and binary orbital periods, respectively. These unambiguously identify \IGRJ\ as a magnetic CV of the Intermediate Polar (IP) type. The energy resolved light curves at both periods reveal amplitudes decreasing with increasing energy, with the orbital modulation reaching  $\sim100\%$ in the softest band.
The energy spectrum shows optically thin thermal emission with an excess at the iron complex, absorbed by two dense media 
(${\rm N_{H}\sim10^{22-23}\,cm^{-2}}$), partially covering the X-ray source.
These are likely localised in the magnetically confined accretion flow above the WD surface and at the disc rim, producing the energy dependent spin 
and orbital variabilities, respectively.  
\IGRJ, joins the group of strongest orbitally modulated IPs now counting 
four systems. 
Drawing similarities with low-mass X-ray binaries displaying orbital dips, 
these IPs should be seen at large orbital inclinations allowing azimuthally 
extended absorbing material fixed in the binary frame to intercept the line of sight. For \IGRJ, we estimate ($50^o\,\lesssim\,i\,\lesssim\,70^o$). Whether also the mass accretion rate plays a role in the large orbital modulations in IPs cannot be established with the present data.

\end{abstract}


\section{Introduction}

Our understanding of the hard X-ray sky considerably improved 
thanks to the deep surveys carried out 
by the \Swift/BAT and \INT/IBIS satellites 
at energy greater than 20 keV \citep{cusumano10,bird16}. About 20 per cent of the galactic sources 
detected  in these surveys are CVs, among which the majority host magnetic WD primaries (MCVs). 
These are divided in two subclasses, depending on the WD magnetic field strength and 
degree of asynchronism. Polars possess stronger magnetic fields (B$\gtrsim10^{7}$ G),  which 
eventually synchronise the binary system (${\rm P}_{\rm spin=\omega} \sim {\rm P}_{\rm orb=\Omega}$),
and thus do not possess an accretion disc. The Intermediate Polars (IPs) instead
harbour asynchronously rotating WDs 
\citep[${0.01\lesssim{\rm P}_{\rm \omega}/{\rm P}_{\rm \Omega}<1}$; see e.g.][]{bernardini17} 
and  consequently are believed to possess weaker magnetic fields (B$\leq10^{6}$ G), and thus a truncated disc at the magnetospheric radius may form.  
For two recent reviews on magnetic WDs and CVs, respectively, see \cite{ferrario15} and \cite{mukai17}.   
   
Optical follow ups of the still unidentified sources in the BAT and IBIS catalogues provide suitable MCVs candidates \citep[see e.g.][and references therein]{masetti13,halpern15}. 
However, a proper classification resides in the X-rays and in particular in the detection of a coherent signal at the WD spin period and in the characterization of the broad-band energy spectrum \citep[see e.g.][and references therein]{bernardini12,bernardini14}. Short X-ray periodicities imply 
that the inner accretion flow follows the WD magnetic field lines, finally reaching the compact object surface, testifying that the WD is indeed magnetic. Since the flow has a supersonic velocity, a standoff shock forms and matter in the post shock region (PSR) cools (and slows) down via 
bremsstrahlung (hard X-ray) and cyclotron (optical/nIR) radiation \citep{aizu73,wu94,cropper99}, the efficiency of which mainly depends on the magnetic field intensity \citep{woelk96,fischer01}. 
Cyclotron is more efficient in high intensity magnetic field systems, e.g. Polars, while IPs are bremsstrahlung dominated systems and so, in general, harder X-ray emitters. MCV X-ray spectra are also characterised by the ubiquitous presence of a Fe K$_{\alpha}$ line at 
6.4 keV, due to Compton reflection from the nearly neutral WD surface \citep{mukai17} and, 
in some cases, by a soft ($\sim20-100$ eV) X-ray blackbody emission due to thermalisation of 
the hard X-rays, likely from the WD surface polar region. Nowadays, thanks to high S/N X-ray 
instruments like \XMM, the blackbody soft component is frequently detected 
also in IPs and not only in Polars as it 
originally seemed to be the case \citep[see e.g.][and reference therein]{bernardini17}.

\IGRJ\ is one of the still unclassified BAT and IBIS sources, for which X-ray 
properties were not studied yet. It was proposed to be a MCV, due to its 
optical spectroscopic characteristics, displaying strong emissions of Balmer, He
\,I and He\,II \citep{masetti13}. We here present the first simultaneous X-ray 
and optical data collected with \XMM\ complemented with \Swift/BAT high-energy spectral coverage that allow us to 
unambiguously identify it as a new member of the IP class with a particularly strong X-ray orbital modulation.

\section{Observations and data reduction} 
\label{sec:obs}

\subsection{\textit{XMM-Newton} observations}

\IGRJ\ was observed on 2017-01-20 by \XMM\ with 
the European Photo Imaging Cameras \citep[EPIC: PN, MOS1 and MOS2][]{struder01,turner01,denherder01} 
as main instruments, complemented with simultaneous optical monitor \citep[OM,][]{mason01} photometry. 
The observation details are reported in Table \ref{tab:observ}. Data were processed using the 
Science Analysis Software (\textsc{SAS}) version 16.1.0 and the latest calibration files available 
in 2017 October.

\begin{table*}
\caption{Summary of main observation parameters for all instruments. Uncertainties are at $1\sigma$ confidence level.}
\begin{center}
\begin{tabular}{cccccccc}
\hline 
Source Name            & Telescope            & OBSID        & Instrument    & Date        & UT$_{\rm start}$ & T$_{\rm exp}$ $^a$ & Net Source Count Rate\\
Coordinates (J2000)$^{b}$&               &              &        & yyyy-mm-dd      & hh:mm & ks      &    c/s                  \\
\hline
IGR J14257-6117    & \XMM\                &  0780700101  & EPIC-PN$^c$   & 2017-01-20  & 07:02 & 37.5 & $0.475\pm0.006$ \\
                   &                      &              & EPIC-MOS1$^c$ & 2017-01-20  & 06:57 & 37.7 & $0.152\pm0.002$     \\ 
RA=14:25:07.58	   &                      &              & EPIC-MOS2$^c$ & 2017-01-20  & 06:57 & 37.7 & $0.160\pm0.002$     \\               
Dec=-61:18:57.8    &                      &              & OM-V$^d$      & 2017-01-20  & 07:03 & 33.3 & $18.62\pm0.13^e$    \\
                   & \Swift\              &              & BAT$^f$       &             &	      & 8581 & $1.9\pm0.3\times 10^{-5}$ \\ 
\hline              
\end{tabular}
\label{tab:observ}
\end{center}
\begin{flushleft}
$^a$ Net exposure time.\\
$^b$ Coordinates of the optical counterpart. \\
$^c$ Small window mode (thin filter applied). \\
$^d$ Fast window mode. The central wavelength of the V filter is 5430 \AA. \\ 
$^e$ OM instrumental magnitude.\\
$^f$ All available pointings collected from 2004 December to 2010 September are summed together. \\
\end{flushleft}
\end{table*}


Source photon event lists and spectra for EPIC cameras were extracted 
from a circular region of radius 40 arcsec. 
The background was extracted in the same CCD where the target lies, selecting a region free from sources contamination, avoiding CCD gaps.
The observation was affected by moderate  particle background epochs that were conservatively removed in  all instruments for the spectral analysis, while for the timing analysis, the whole dataset was used.

Background-subtracted PN and MOSs light curves were produced with the task \textsc{epiclccorr} in several 
energy bands, with different bin size depending on the source and background rates.
The event arrival times were barycentered by using the 
task \textsc{barycen}. Before fitting, spectra were rebinned using \textsc{specgroup}. A minimum of 
50 and 25 counts in each bin for PN and MOSs, respectively, and a maximum oversampling of the energy 
resolution by a factor of three were set. Phase resolved spectra were also extracted at the minimum and maximum of the spin and orbital cycle. 
The response matrix and the ancillary files were generated using the tasks 
\textsc{rmfgen} and \textsc{arfgen}, respectively. The RGS1 and RGS2 spectra were of poor S/N for a useful analysis. PN and MOSs spectra were fitted together by using \textsc{Xspec} version 12.9.1p package \citep{arnaud96}.

The OM was operated in fast window mode using the
V-band (5100--5800 \AA) filter. The background subtracted light curve was generated with the 
task \textsc{omfchain} with a bin time of 10 s and then the barycentric correction was applied. 

\subsection{The \Swift\ observations}

The \Swift/BAT eight-channel spectra and response file from the first 66 months of BAT monitoring \citep{baumgartner13} were downloaded from the publicly available archive at the Palermo BAT website\footnote{http://bat.ifc.inaf.it/}. \IGRJ\ is detected up to 80 keV, to which we restricted the spectral analysis.

\section{Data analysis and results}
\label{sec:danalysis}

\subsection{Timing analysis} 

\begin{figure*}
\begin{center}
\begin{tabular}{cc}
\includegraphics[angle=270,width=3.3in]{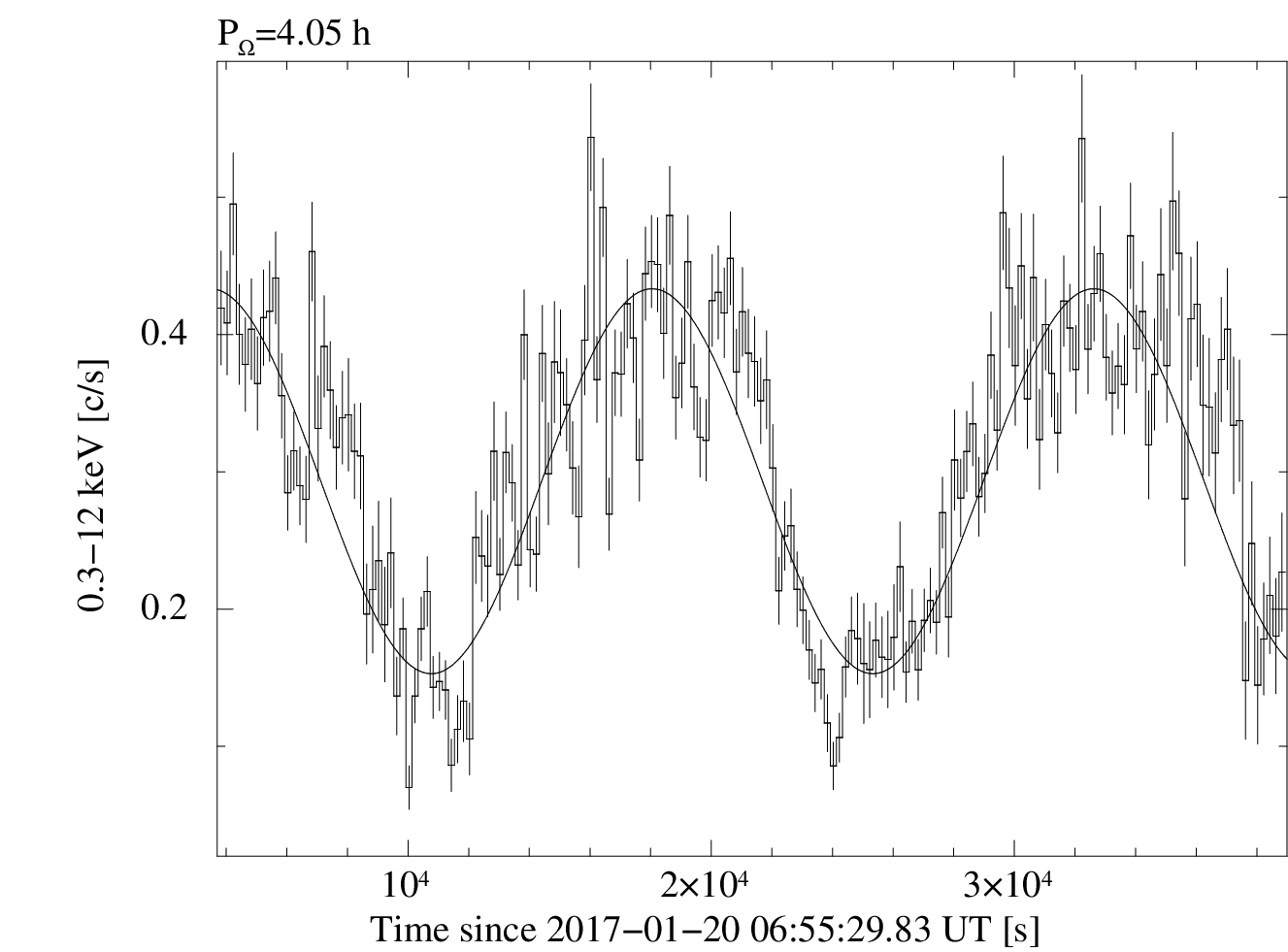} & \includegraphics[angle=270,width=3.3in]{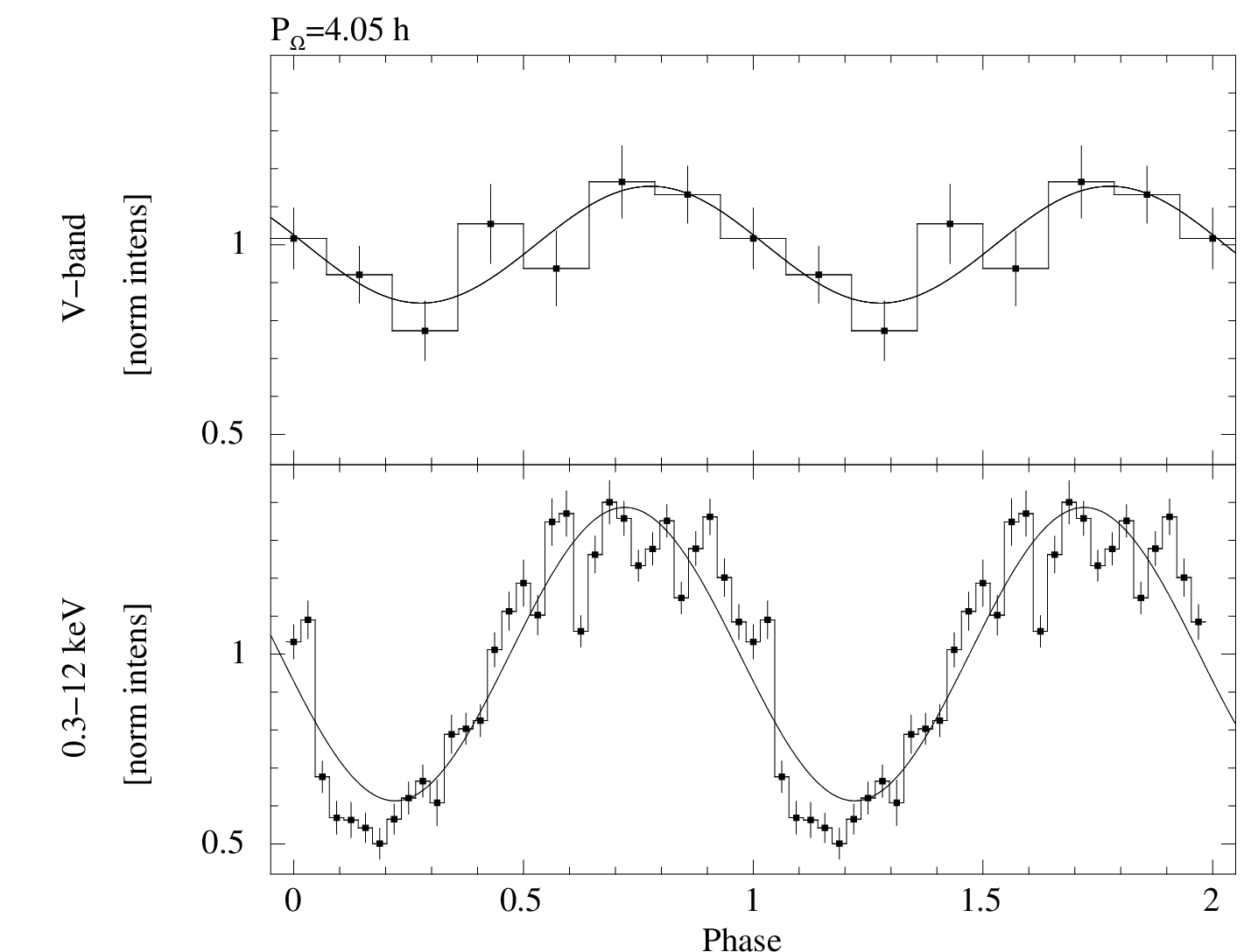}\\ 
\end{tabular}
\caption{\textit{Left:} PN plus MOSs 0.3--12 keV background subtracted light curve of \IGRJ.  
Short-term, the WD spin, and long-term, the orbital,  modulations are present. 
\textit{Right:} Background subtracted V-band (top) and 0.3--12 keV (bottom) light curves, folded at 
the orbital period. Two cycles are shown for plotting purposes. 
The reference folding time is the integer of the observation start time.
In both panels, the solid line represents a sinusoid at the 4.05 h period.} 
\label{fig:orbital}
\end{center}
\end{figure*}

\begin{figure}
\begin{center}
\includegraphics[angle=270,width=3.3in]{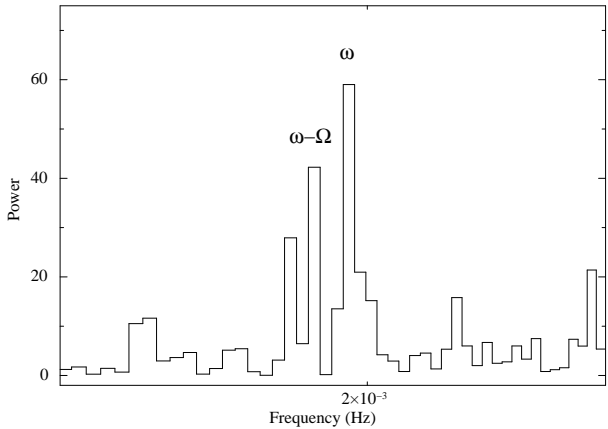} 
\caption{PN 0.3--12 power spectrum zoomed in the spin ($\omega$) and beat ($\omega-\Omega$) frequency region.}
\label{fig:pszoom}
\end{center}
\end{figure}

The results presented in this section refer to the analysis of the sum of the source event files and 
background subtracted light curves of the three EPIC cameras. 
Two and a half cycle of a long-term periodic variability are clearly present in the background 
subtracted 0.3--12 keV light curve with superposed periodic short-term variations 
(Figure \ref{fig:orbital}). The X-ray light curve was heavily rebinned to wash out any short-term variation. A fit with a single sinusoid plus constant  gives a period of 4.05$\pm$0.06 h and a pulsed fraction (PF)\footnote{$\rm PF=(F_{max}-F_{min})/(F_{max}+F_{min})$, where $\rm F_{max}$ and $\rm F_{min}$ are respectively the maximum and minimum fluxes of the sinusoid at the fundamental frequency.} of 44$\pm2$ per cent. We naturally interpreted it as the binary system orbital period (P$^{X,lc}_{\Omega}$). All 
uncertainty are hereafter reported at $1\sigma$ confidence level. The fit slightly improves including the first harmonic, giving a period of 4.02$\pm$0.04 h.
Then, the power spectra of the 0.3--12 keV source event file was computed, 
which showed besides 
the strong low frequency peak due to the orbital modulation, two less intense and close-by 
higher frequency ($\sim 2$ mHz) peaks (Figure \ref{fig:pszoom}). 
A phase-fitting technique \citep[see e.g.][]{dallosso03} was used to accurately 
determine the period of the stronger signal. This results 509.5$\pm$0.5 s 
with PF$=9.7\pm1.6$ per cent. A weaker peak at 
its first harmonic is also detected at 7$\sigma$ confidence level, with a
PF of 3.6$\pm1.6$ per cent (see Table \ref{tab:pf}).
We interpreted the 509.5\,s period as the spin period of the accreting primary (P$^X_{\omega}$). 
Since such slow rotators are only encountered in high mass X-ray binaries and never in LMXBs, and
since the optical characteristics of \IGRJ\ are typical of CVs/LMXBs, we can safely identify it
as a magnetic CV of the IP type. Its spin-to-orbit period ratio of 0.03 
locates \IGRJ\ among the majority
of asynchronous systems confirmed so far in the spin-orbital period plane \citep{bernardini17}.
The second weaker signal is instead found at the slightly longer period of 
526.8$\pm$2.0 s, by using the \textsc{FTOOLS}\footnote{http://heasarc.gsfc.nasa.gov/ftools/} task 
\textsc{efsearch} \citep[][]{blackburn95}. This is interpreted as 
the sideband ($\omega-\Omega$, the beat) between the spin and the orbital periods. We note that this is the stronger sideband usually found in IPs.
The spin-to-beat amplitude ratio is found to be 1.4$\pm$0.2
(see also Sect.  \ref{sec:disc}). From the latter, we also derive  
P$^{X,side}_{\Omega}=4.30\pm0.53$ h, which is
consistent within $1\sigma$
with the orbital period obtained from fitting the X-ray light curve. For a summary of the timing properties of \IGRJ\ see Table \ref{tab:time}. 

The V-band light curve instead did not show obvious periodic variability, 
but when folded at the X-ray orbital period a modulation is detected above
3$\sigma$  confidence level. It is much weaker than that in the X-rays, its PF being only 
15 per cent (Figure \ref{fig:orbital}). 

\begin{figure*}
\begin{center}
\begin{tabular}{cc}
\includegraphics[angle=270,width=3.3in]{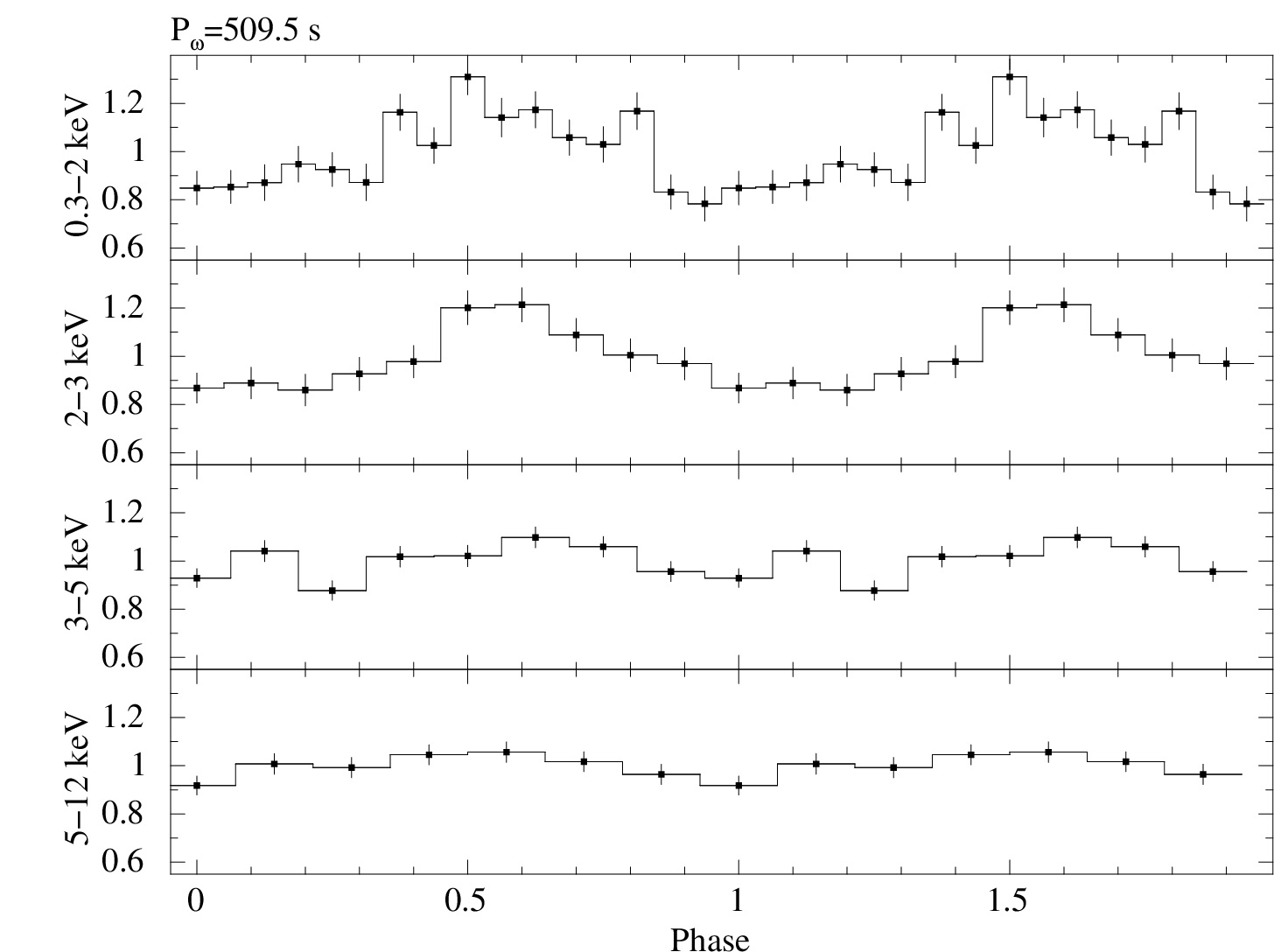} & \includegraphics[angle=270,width=3.3in]{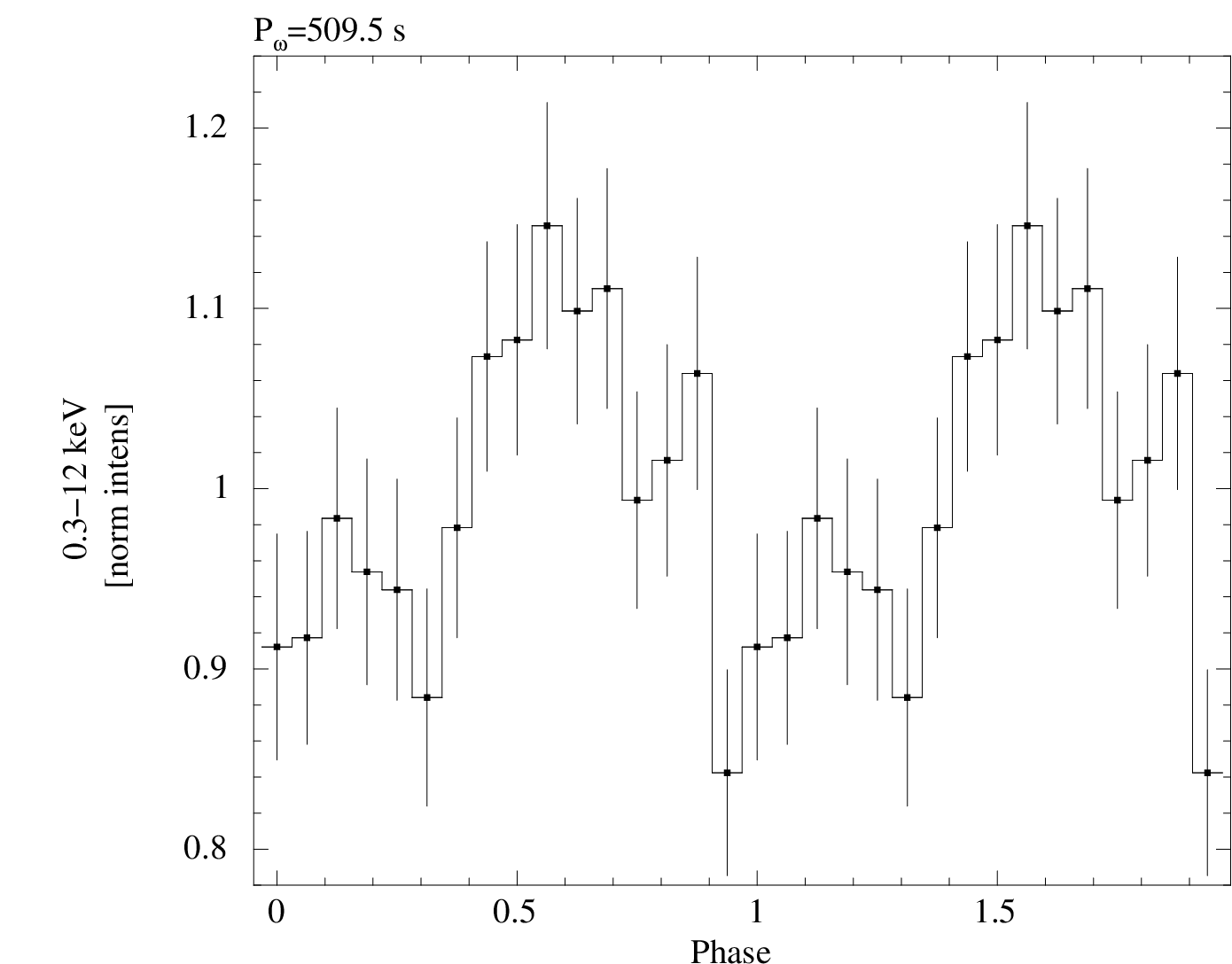}\\ 
\end{tabular}
\caption{\textit{Left:} X-ray (PN plus MOSs) normalised spin-folded light curves in different energy intervals. 
Energy increases \textit{from top to bottom}. Two spin cycles are shown for plotting purposes. 
The reference folding time is the integer of the observation starting time. 
The PF decreases as the energy increases (cfr Table \ref{tab:pf}). 
\textit{Right}: The spin light curve in the 0.3--12 keV band also 
reveals the first harmonic (see text).}
\label{fig:pulseprof}
\end{center}
\end{figure*}

To inspect spectral changes along the spin period, the background subtracted light curves were folded at P$^X_{\omega}$  and the hardness 
ratios (HRs, defined as the count rate ratio in each 
phase bins between two selected energy ranges) were then computed. Spectral hardening at spin minimum ($\sim\phi=0.0-0.4$) was detected. 
To quantify the variation of the spin 
signal with respect to the energy interval, the PF was computed in five energy bands (0.3--1, 1--3, 
3--5, and 5--12 keV) by fitting the modulation with a sinusoid at the fundamental frequency. 
The PF slightly decreases when the energy increases from a maximum 17.5 per cent 
(0.3--1 keV) to a minimum of 6.2 percent (5--12 kev; Table \ref{tab:pf} and Figure 
\ref{fig:pulseprof}). This behaviour is indicative of
photoelectric absorption from neutral material localised above the X-ray emitting polar region.

\begin{figure}
\begin{center}
\begin{tabular}{cc}
\includegraphics[angle=270,width=3.3in]{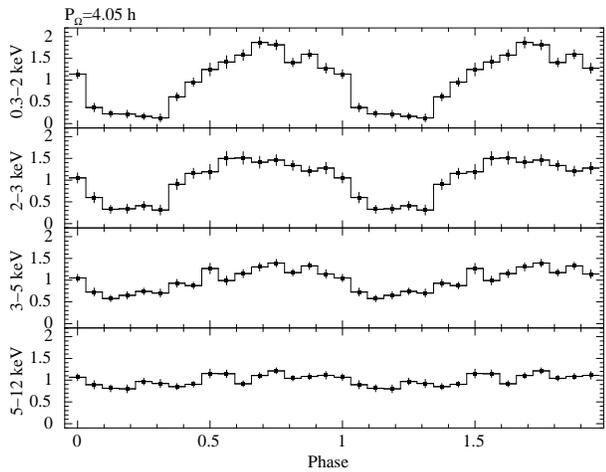} \\ 
\end{tabular}
\caption{X-ray (PN plus MOSs) normalised 
orbital modulations in different energy intervals. 
Energy increases \textit{from top to bottom}. Two cycles are shown for plotting purposes. 
The  reference folding time is the integer of the observation starting time 
(Table \ref{tab:observ}). 
The X-ray PF decreases as the energy increases (Table \ref{tab:pf}). 
}
\label{fig:orbprof}
\end{center}
\end{figure}

Spectral variations (HRs) were also inspected as a function of the orbital  period.
The amplitude of the orbital modulation also decreases with increasing energy with PF of
$84$ per cent in the 0.3-1\,keV to $\sim12$ per cent in the hardest band (Table \ref{tab:pf} and Figure
\ref{fig:orbprof}).
The spectrum is clearly harder during orbital minima, indicating the presence of additional neutral material localized at a fixed region in the binary frame.

\begin{table}
\caption{Timing properties of \IGRJ. Uncertainties are at $1\sigma$ confidence level.  
From left to right: P$^X_{\omega}$ (X-ray spin period);
 P$^X_{side}$ (X-ray sidebands); P$^{X,side}_{\Omega}$ (orbital period derived from 
X-ray sidebands); 
P$^{X,lc}_{\Omega}$ (orbital period derived from X-ray light curve fitting); 
A$^X_{\omega}$/A$^X_{side}$ (spin to sideband X-ray amplitude ratio);
P$^{A}_{\Omega}$(adopted orbital period in this work). 
}

{\small
\begin{center}
\tabcolsep=0.1cm
\begin{tabular}{cccccc}
\hline 
               &                &                       &                     &                             &   \\
P$^X_{\omega}$ &  P$^X_{side}$  & P$^{X,side}_{\Omega}$ & P$^{X,lc}_{\Omega}$ & A$^X_{\omega}$/A$^X_{Side}$ &  \textbf{P$^{A}_{\Omega}$}\\
 s & s  & h & h &   & h  \\
\hline 
\\
509.5$\pm$0.5 & 526.8$\pm$2.0  & 4.30$\pm$0.53   & 4.05$\pm$0.06 & 1.4$\pm$0.2 & 4.05$\pm$0.06\\
\hline 
\end{tabular}
\label{tab:time}
\end{center}}
\begin{flushleft}
\end{flushleft}
\end{table}

\begin{table}
\begin{center}
\tabcolsep=0.1cm
\caption{Pulsed fraction vs energy. Results refer to the fundamental frequency ($\omega$, and $\Omega$, Table \ref{tab:time}, column 2 and 4). Uncertainties are at $1\sigma$ confidence level.}
\begin{tabular}{cccccccc}
\hline     
\multicolumn{6}{c}{Pulsed Fraction} \\   
Period  & 0.3--2 keV& 2--3 keV  & 3--5 keV  & 5--12 keV   & 0.3--12 keV$^a$ \\
\% & \%   &  \%  & \%   & \%      &  \%   \\
\hline 
\\
$\rm P_{\omega}$ &  18$\pm$2 & 15$\pm$3     & 7$\pm$2     & 6$\pm$2    & 9.7$\pm$1.6   \\
\hline 
$\rm P_{\Omega}$ & 84$\pm$3    & 60$\pm$3 & 33$\pm$2  & 12$\pm$2     & 44$\pm$2 \\      
\hline
\end{tabular}
\label{tab:pf}
\end{center}
\begin{flushleft}
$^a$ The first harmonics have $\rm PF_{2\omega}=3.6\pm1.6$ per cent and 
$\rm PF_{2\Omega}=8\pm2$ per cent. \\
\end{flushleft}
\end{table}

\subsection{Spectral analysis}
\label{sec:spec}

Fits of the broad-band average spectrum were made on the three EPIC cameras plus BAT spectra 
simultaneously, encompassing the range from 0.3 to 80 keV. An inter-calibration constant 
(fixed to one for the PN only) was used to account for instrument calibration discrepancies and 
spectral variability due to the fact that the BAT data are not simultaneous. All model parameters 
were linked between different instruments with the exception of the multiplicative constants.

\IGRJ\ has a thermal spectrum showing emission at the iron complex. These are characteristics 
generally observed in CVs and particularly in the magnetic systems, which usually have 
multi-temperature spectra locally absorbed by dense cold material 
\citep[see e.g.][]{Done95,EzukaIshida99,beardmore00,
demartino04,bernardini12,bernardini13,mukai15,bernardini17}. 
Consequently, the broad-band spectrum 
was fitted using a model consisting of 
an optically thin plasma component (\textsc{mekal} 
or \textsc{cemekl} in \textsc{Xspec}), with metal abundances (A$_{\rm Z}$) with respect to Solar\footnote{We set the abundance to that of the ISM from \cite{wilms00}}  left free to 
vary, plus a narrow Gaussian line 
kept fix at 6.4 keV accounting for the fluorescent Fe K$_{\alpha}$ feature, 
all absorbed by a total (\textsc{phabs}) and two partial (\textsc{pcfabs}) covering 
 columns. Indeed, a fit with only one absorbing partial covering column shows clear residuals below 1 keV. Therefore, as is the case for other IPs showing a 
strong orbital modulation \citep{bernardini17}, a second absorber was included in the 
spectral fit. The use of two \textsc{pcfabs} is also justified by the fact that the 
intensities of the spin and orbital modulations decrease as the energy increases  
and thus they are likely due to different components. 
We obtained statistically acceptable fits using both a multi-temperature plasma  
(\textsc{cemekl}) ($\chi^{2}_{\nu}=0.98$, 319 d.o.f.) 
and a single temperature plasma (\textsc{mekal})  
($\chi^{2}_{\nu}=1.03$, 319 d.o.f.) (Table \ref{tab:averagespec}). 
 
The total absorber $\rm N_{phabs}=2.2-3.6\times 10^{21}\,cm^{-2}$ is 
one order of magnitude lower than that of the ISM in the direction of the source \citep{kalberla05}. The partial covering absorbers have densities of the order of $2-3\times10^{22}$ cm$^{2}$ (Pcf1) and $2-3\times10^{23}$ cm$^{-2}$ (Pcf2) and their covering fractions are large: $\sim80$ per cent (Pcf1) and $\sim 65$ per cent (Pcf2). 
The spectrum does not require a soft optically thick component 
(e.g. kT$_{\rm BB}\sim20-100$ eV). In the case of the \textsc{cemekl}, the maximum plasma temperature is 
poorly constrained even not fixing to 1 the power-law $\alpha$ index and we derived 
a $3\sigma$ lower limit of 35 keV.  This is much higher than that 
derived using \textsc{mekal}, where kT$=18\pm^{5}_{2}$ keV. However, we note that in the latter 
case the temperature represents an average over the entire PSR, 
so the two temperatures are not expected to be consistent. 
We note that even if a lower limit, the \textsc{cemekl} temperature should be considered as a more 
reliable estimate of the shock temperature. We also checked whether a reflection component is required in the spectral fits, as indicated by the presence of the 6.4\,keV iron fluorescent line (EW=180$\pm$20eV), that would alleviate the problem of such high lower limit to the maximum temperature. Such component is however not statistically required in the spectral fits. 

To obtain an estimate of the mass of the accreting WD, the broad-band 
continuum spectrum was also fitted (above 3 keV only) with the more physical model developed by \cite{suleimanov05}, which takes into account both temperature and gravity gradients within the PSR. This gives a loosely constrained mass: M$_{\rm WD}=\rm0.58\pm0.20 \, 
M_{\odot}$ ($\chi^{2}_{\nu}=1.13$, 189 d.o.f.).  It is however consistent 
within $\sim1\sigma$ with the lower limit to the mass derived using the maximum 
\textsc{cemekl} temperature (${\rm M_{\rm WD}\geq0.78\,M_{\odot}}$). Clearly the \Swift/BAT
spectrum is of too low quality to obtain a precise WD mass and 
higher S/N data are needed \citep[see][]{suleimanov16,shaw18}.

To investigate the role of spectral parameters in generating the X-ray spin modulation, a spin-phase resolved spectroscopic (PPS) analysis was performed. EPIC spectra extracted at spin maximum and minimum were fitted separately using both models presented in Table \ref{tab:averagespec}. N$_{\rm H_{Ph}}$, A$_{\rm Z}$, and kT (which is otherwise unconstrained) were fixed at their average spectrum best-fitting values. All other parameters, hence the two partial covering absorbers and the Gaussian 
normalization, were left free to vary. Likely due to the low PF 
(Table \ref{tab:pf}), all free parameters are constant within 
less than $2\sigma$. A similar analysis was also performed on the EPIC spectra 
extracted at orbital maximum and minimum. This time, only the two partial covering 
components were left free to vary and all other model components were fixed at their 
average spectrum best-fitting values. The spectrum at orbital minimum is clearly harder. 
This is due to a significant change in both partial covering absorbers 
that increase at orbital minimum. In particular, in the case of 
both \textsc{cemekl} and \textsc{mekal}, N$_{\rm H_{Pc1}}$ and N$_{\rm H_{Pc2}}$ 
increase by a factor of $\sim3$ and $\sim2$, respectively. On the other hand, 
the covering fraction of Pc1 significantly increases at orbital minimum, but not that of the higher density 
absorber (Pc2),  which is found to be
constant within $3\sigma$ (Table \ref{tab:pps_orb}). Therefore, while we are unable to
assess which of the two absorber is mainly responsible for the spin variability, 
the lower density one (Pc1) appears to be the major contributor to the orbital variability.

\begin{table*}
\caption{Parameters of the best fit models to the averaged broad-band spectrum. 
The absorbed 0.3--10 keV and unabsorbed bolometric 
(0.01--200 keV) fluxes are reported in the last two columns. 
Uncertainties are at $1\sigma$ confidence level.}
{\small
\begin{center}
\tabcolsep=0.18cm
\begin{tabular}{ccccccccccccc}
\hline 
\\
mod. & N$_{\rm H_{Ph}}$  & N$_{\rm H_{Pc1}}$  &  cvf & N$_{\rm H_{Pc2}}$ &  cvf  & kT & n & A$_{\rm Z}$  & EW  &  F$_{0.3-10}$ &   F$_{\rm X,bol}$ &  $\chi^2$/dof  \\
     & $10^{22}$         & $10^{22}$          &      &    $10^{22}$      &  &                   &  $10^{-3}$  &             &       & $10^{-12}$  &    $10^{-12}$   &   \\
 & cm$^{-2}$ & cm$^{-2}$          & \%  &  cm$^{-2}$         & \%   & keV          &              &              & keV & erg/cm$^2$/s & erg/cm$^2$/s & \\         
\hline 
\\
cemek$^a$ & 0.22$\pm$0.10 & 1.87$\pm_{0.23}^{0.32}$ & 81$\pm$3 & 20$\pm$2 & 65$\pm$4 & $>35$ $^b$ & 10.0$\pm$0.5  & 1.3$\pm$0.3 & 0.18$\pm$0.02 & 4.6$\pm$0.1 & $\sim19.7$ & 0.98/319 \\
mek & 0.36$\pm$0.07 & 3.1$\pm$0.8 & 74$\pm$3 & 28$\pm$6 & 63$\pm$3 & $18\pm_{2}^{5}$ & 5.3$\pm$0.4  & 0.68$\pm$0.13  & 0.18$\pm$0.01& 4.6$\pm$0.1 & $\sim21.1$ & 1.03/319 \\
\hline
\end{tabular}  
\label{tab:averagespec}                      
\end{center}} 
\begin{flushleft}
$^a$ Multi-temperature power-law index $\alpha$ fixed to 1.\\
$^b$ $3\sigma$ lower limit. The best fitting value is 80 keV, but poorly constrained.\\ 
\end{flushleft}                               
\end{table*}

\begin{figure}
\begin{center}
\includegraphics[angle=270,width=3.3in]{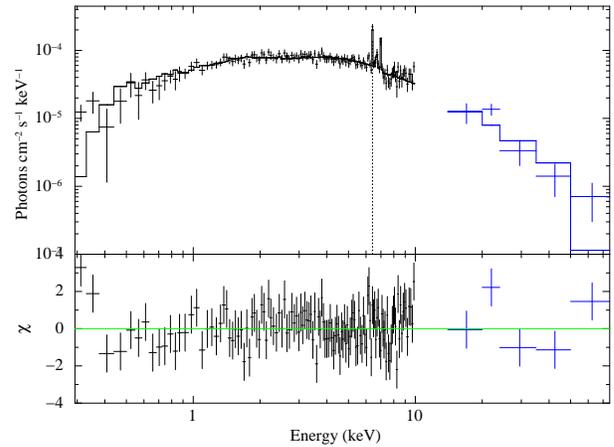} 
\caption{Broadband unfolded spectrum of \IGRJ. Post fit residuals are shown in the \textit{lower panel}. Black points are \XMM/PN data (0.3--10 keV), blue points are \Swift/BAT data (15--80 keV). The fit is made simultaneously on all EPIC cameras, but for sake of readability, only the PN data are shown. The dotted line mark the 6.4 keV Gaussian component, while the solid line is the composite model.}
\label{fig:avrspec}
\end{center}
\end{figure}

\begin{table}
\caption{Spectral parameters at maximum (Max) and minimum (Min) of the orbital modulation (Orb.). All other parameters are fixed to their average spectrum best-fit values. Uncertainties are at $1\sigma$ confidence level.}
{\small
\begin{center}
\tabcolsep=0.05cm
\begin{tabular}{cccccccc}
\hline 
\\
model  & Orb.  &  N$_{\rm H_{ Pc1}}$    & cvf  & N$_{\rm H_{ Pc2}}$   & cvf & F$_{0.3-10}$ & $\chi^2$/dof    \\
       &       &  10$^{22}$ cm$^{-2}$  &  \%  & 10$^{22}$ cm$^{-2}$ & \%  &   $10^{-12}$ &   \\
       &       &                       &      &                     &     &   \ergscm    &   \\ 
\hline  
cemek & Max  & 1.5$\pm$0.2  & 80$\pm$1 &  15.2$\pm$1.2 &  58$\pm$2 & 5.0$\pm$0.1  & 1.07/275 \\                                                                                                          
      & Min  & 4.6$\pm$0.8  & 92$\pm$1 &  35$\pm$4     &  71$\pm$5 & 3.4$\pm$0.1  & 1.04/127  \\
\hline 
mek   & Max  & 2.4$\pm$0.3  & 70$\pm$2 &  25$\pm$2 &  60$\pm$2 & 5.1$\pm$0.1  & 1.10/275 \\                                                                                                          
      & Min  & 6.4$\pm$1.0  & 91$\pm$1 &  48$\pm$5 &  71$\pm$4 & 3.4$\pm$0.1  & 1.08/127  \\
\hline
\end{tabular} 
\label{tab:pps_orb}
\end{center}}
\end{table}

\section{Discussion and conclusions}
\label{sec:disc}

\IGRJ\ lies very close to the galactic plane ($b\sim-0.5^{o}$). The total galactic absorption 
in the source direction is high \citep[$1.6\times10^{22}$ cm$^{-2}$][]{kalberla05}. However, 
N$_{\rm H}$ derived from spectral fits is a factor of $\sim4-7$ lower than that, pointing toward 
a close-by galactic object \citep[see also][for a similar conclusion derived from V-band 
absorption]{masetti13}. To get an estimate of the source distance we use near-IR data. 
\IGRJ\ is listed in the 2MASS catalog as 2MASS J14250758-6118578. It is only detected in the 
J-band, with J$=16.194\pm0.096$ mag (upper limits of H$>$14.8 and K$>$14.4 mag). 
We dereddened it assuming 
N$_{\rm H}=2.9\times10^{21}$ cm$^{-2}$ (as derived from the average X-ray 
spectrum), which translates \citep{Guver_Ozel09} in A$_{\rm V}=1.3$ (and A$_{\rm J}=0.34$). 
For a CV in a 4.05 h orbit, a M3.5 donor with M$_{J}=7.1$ mag is expected \citep{knigge11}. 
Assuming that the donor is totally contributing to the J-band 
flux (when dereddened, J=15.85 mag) we derive a distance 
$\rm d=563\pm225$ pc (assuming a 40 per cent uncertainty). 
Adopting this distance, we estimate the accretion luminosity as:
${\rm L_{acc}=GM\dot{M}/R\sim L_{X,bol}\sim7.6\times10^{32}}$ erg/s, , where ${\rm L_{X,bol}}$ is evaluated over a the wide range 0.01--100 keV. Adopting a conservative 
lower limit to the WD mass of $\rm 0.58\,M_{\odot}$, this translates into an 
upper limit to the mass accretion rate, $\rm \dot M\lesssim 
1.4\times10^{-10}\,M_{\odot}\,yr^{-1}$. If the distance is much larger, as 
from the upper envelope of the distance estimate ($\sim$900\,pc), the mass accretion rate is bounded
to an upper limit of $\rm \lesssim 4\times10^{-10}\,M_{\odot}\,yr^{-1}$. 
As in the case of the majority of IPs 
\citep[e.g. see][]{bernardini12,bernardini17}, $\rm \dot M$ is found to be lower 
than the secular mass transfer rate predicted by models of the present 
day CV population, in the case of a binary above the orbital period gap evolving 
through magnetic braking in a 4-h orbit 
\citep[$\rm \dot M\sim5\times10^{-9}\,M_{\odot}\,yr^{-1}$;][]{howell01}. 
These estimates should be considered with caution because a non-negligible fraction of the X-ray emission could be reprocessed in the accretion flow (e.g. the magnetically confined accretion flow above the shock and 
the accretion disc) and radiated at lower energies. Low
mass transfer rates have also been found in an increasing number of CVs above the 2-3\,h
orbital period gap using the effective temperature of the unheated primary, when detected
 \citep[see][]{pala17}. 
 
\IGRJ\ is then an IP above the gap with a spin-to-orbit period ratio 
${\rm P}_{\omega}/{\rm P}_{\Omega}\sim0.03$, and its position in the 
P$_{\omega}$ vs P$_{\Omega}$ plane \citep[see Figure 6, left panel in][]{bernardini17} 
falls where the majority of the systems of its class lie, confirming that the 
observed present
day population of IPs is dominated by systems above the period gap. Whether the 
paucity of IP systems below the gap \citep[see also][]{pretorius14}, 
is due to selection effects in their discovery 
(faint X-ray sources) or indeed almost all IPs evolve into low-field Polars, still 
remains an open problem to be addressed with future sensitive X-ray survey missions such as
{\it eROSITA}. 

\IGRJ\ has been found to show the spin, the beat and orbital modulations. 
The spin variability is found to be much weaker than that at the orbital period. The energy dependence of the rotational
modulation is a characteristic also found in the majority of IP systems and consistent with the
accretion curtain scenario \citep{rosen88} where the magnetically confined accretion occurs
in an curtain-shaped flow. The hardening at spin minimum is due to the curtain pointing towards
the observer when the absorption in the pre-shock flow is maximum. The spin variability indicates
that matter is accreted via a disc. However, the detection of a non-negligible
variability at the $\omega-\Omega$ sideband implies that material is also overflowing the disc
at a fraction of $\sim45$ per cent of the total flow. Such hybrid accretion geometry is also observed
in many other IPs,  indicating it is not so uncommon in these systems
\citep{bernardini12,hellier14,bernardini17}.
 
\IGRJ\ is also one of the members of the IP class that shows an X-ray light curve 
strongly modulated at the orbital period.  Energy dependent 
X-ray orbital modulations appear to be a common property of IPs and, in some systems, also 
found to vary on timescales of years \citep[see e.g.][]{parker05,bernardini17}, with the
unique case of FO\,Aqr that entered in a low state in spring 2016 and recovered its
high state at the end of 2016 \citep{kennedy16,kennedy17}. 
Long-term changes of the amplitudes of spin, orbital and beat variabilities were also observed in IPs at both X-rays \citep{norton97,beardmore98,staude08} and  UV/optical wavelengths
\citep{demartino99}, accompanied by moderate changes in their brightness. Such changes are ascribed to variations in the mass accretion rate, which in turn modifies the accretion geometry. In particular changes in the spin-to-beat amplitudes are interpreted
as an increase or decrease of the disc-overflow contribution \citep[see][]{hellier14}. 
The presence of a strong orbital and energy
dependent modulation indicates absorbing material fixed in the orbital frame. The lack of
a spectroscopic orbital ephemeris does not allow us to correctly locate the 
superior conjunction of the donor or the WD but, drawing similarities with other IPs
also showing energy dependent orbital modulations, this material should be  located at the outer disk rim, where the stream of material from 
the companion impacts the disc. The presence of a disc overflow further supports the scenario of an azimuthally extended region.  
\noindent Disc structures are also found in LMXBs seen at relatively high inclinations, the 
so-called "dippers", which show periodic dips at the orbital period,  
generally attributed to partial obscuration of the X-ray emitting
source by a thickened ionised region of the accretion disc \citep{DiazTrigo06}. Sometimes the occurrence of dips is intermittent, as observed in Aql X-1 in two occasions \citep{galloway16}.
In the ultra-compact binary 4U\,1820-303 an orbital modulation is observed
with amplitude changing with X-ray luminosity, but is not energy dependent  
\citep{zdziarski07}. Such variations have been claimed to arise from changes in the mass
accretion rate due to disc precession in both systems. Similar interpretation of a precessing
disc was given for the IPs showing X-ray orbital modulations with 
amplitude changing with time \citep{parker05,norton&mukai07}.  Whether the changes in mass accretion rate 
are due to disc precession or to variable mass transfer rate from the donor star is however unclear. The X-ray emission in these systems 
would then be viewed through higher density columns at different epochs, along the 
precessing period. The high density material (up to $\rm 10^{23}\,cm^{-2}$) would 
then produce a large orbital modulation, expected to reach $100$ per cent at 1 keV, and a significant modulation up to 10 keV \citep{norton89,parker05}. In \IGRJ\ we have inferred
the presence of two local partial 
covering absorbers with high column densities: $\rm N_{H,pcf1} \sim
2\times 10^{22}\,cm^{-2}$, cfv1$\sim 80$ per cent  and $\rm N_{H,pcf2} \sim 
2\times 10^{23}\,cm^{-2}$, cfv2$\sim 60$ per cent.  With the present data we are 
unable to definitively assess which of them is responsible for the weaker
spin amplitude variability and the large orbital modulation, although the lower 
density complex absorber (pcf1) appears to be the major contributor to the orbital modulation.

We then inspected whether the large amplitude orbital variability in \IGRJ\ fits into a general scheme of IPs with a modulation that increases with the increasing X-ray luminosity and thus mass accretion rate.   
An essential parameter to assess this is the binary inclination, which is not well determined in IPs, but
it is proposed to be in excess of 60$^o$ in all IPs showing X-ray orbital modulations \citep{parker05}. 
We then collected the modulation depths \footnote{It is defined as the peak-to-peak amplitude of the sinusoid used to fit each light curve folded at the orbital period, divided by the maximum flux of the sinusoid. The mean level, amplitude and phase of the sinusoid were left free to vary in the fit.}, as measured from $ASCA$ data in the 0.7--2 keV rage, for 12 IPs. For these systems we also collected the distance estimates and when available their binary inclination\footnote{Koji Mukai Website 
(https://asd.gsfc.nasa.gov/Koji.Mukai/iphome/iphome.html)} (see Sect. \ref{sec:appendix}). 
We enlarged the sample to 17 sources by including five additional 
systems found by us to show orbital modulations: \IGRJ, Swift J0927.7-6945 (henceforth J0927) and Swift J2113.5+5422 (henceforth J2113) \citep{bernardini17}, as well as V709\,Cas and NY\,Lup studied in \cite{mukai15}.  For consistency,  we evaluated the modulation depth as defined in Parker et al. 2005. For all sources in the sample, we computed their unabsorbed bolometric luminosity in the 0.01--100 keV range (see Sect. \ref{sec:appendix}) as a
proxy of the mass accretion rate at the epoch of the observation. For V1223\,Sgr, V405 Aur,
and PQ\,Gem there are two epochs of observations and thus both are included in the analysis.
As shown in Table \ref{tab:orb_mod}, \IGRJ\ and FO\,Aqr (in 1997), have shown
the strongest orbital modulation depth (almost 100 per cent), 
although J0927, BG\,CMi have consistent modulation depths within
their uncertainties.
These are found at relatively high luminosity ($\rm 1-6\times10^{33}\,
erg\,s^{-1}$), but only FO\,Aqr and BG\,CMi are known to be moderately high 
inclination systems. Possibly BG\,CMi,  similarly to FO\,Aqr,  has a 
higher binary inclination
than the lower limit estimated so far. Similar or even higher 
luminosity levels are however 
found in other systems showing weaker (20--60 per cent) orbital modulation depths. 
While distances could strongly affect the derived luminosities, the fact that systems at
low inclinations, such as NY\,Lup, YY\,Dra and V1223 Sgr have weak 
orbital modulation depths may suggest that the binary inclination plays a key role in shaping the orbital modulation in these systems. 
For the three IPs for which two epochs are available, no clear relation with luminosity is found, 
but small changes in depths could indicate slight variations in the azimuthal structure of
the responsible region.
Considering only systems for which the detection of 
an orbital modulation is above 3$\sigma$, we are left with nine IPs for which no 
clear relation with the luminosity, and thus the mass accretion rate, is found\footnote{We do not
attempt to relate directly the orbital depth with the mass accretion rate due to 
the additional uncertainty in the WD mass estimates}.  Unless true 
distances are greatly different, the lack of a statistically significant 
correlation of the orbital modulation depths 
with the luminosity could favour a scenario where the sample, 
although still poor, would be mainly constituted 
by moderate-high binary inclination ($\gtrsim 50-60^o$) systems, 
with the only exception being the peculiar low-inclination 
TX\,Col \citep[see][]{FerrarioWickramasinghe99}. The 
the lack of X-ray eclipses
in \IGRJ, J2113, and J0927 also poses an upper limit to the 
inclination of these systems, $i \lesssim 70^o$.
Future precise parallaxes of the sample
that will soon be available with $Gaia$ DR2 release, 
will allow firmer conclusions. 

We here finally note that the discovery of  
\IGRJ\ as an additional strongly orbital modulated IP makes these 
systems ideal targets to be monitored for possible variations in 
their orbital modulation depth. This would allow  to
assess whether this feature is stable or 
may hint to a precessing disc or changes in the mass accretion rate. 
Further observations, especially optical spectroscopy, will be extremely 
useful to determine the true inclination of this binary. 

\appendix
\section{Orbital modulation depths and bolometric luminosities of the IP sample}
\label{sec:appendix}

To explore a possible relation among the orbital modulation depth, the
bolometric luminosity and the inclination, we collected the sources  
presented in \cite{parker05} that were observed with $ASCA$. We selected $ASCA$ data since, due to the effect of the photoelectric absorption, our analysis concentrates in the soft X-ray bands, where the orbital modulations are higher. Two of these sources were also later observed jointly by \XMM\ and $NuSTAR$, namely V1223 Sgr, and NY Lup, together with V709 Cas \citep{mukai15}. We also included those data in our analysis.
Finally, we add three other IPs that were recently observed by 
\XMM\ to show X-ray orbital modulations, namely J0927 and
J2113 \citep{bernardini17}, plus \IGRJ\ (Table \ref{tab:orb_mod}).
For the above 5 sources, the orbital modulation depths as defined by \cite{parker05} were computed in the same 0.2-7\,keV range, for uniformity.

\noindent In order to estimate the source bolometric (0.01--100 keV) fluxes  
we proceeded as follow.
For those sources observed with $ASCA$, we downloaded the public archival BAT 105-month spectra\footnote{https://swift.gsfc.nasa.gov/results/bs105mon/}. Since
the spectra span a long period of time that does not overlap with the  $ASCA$  
observations,  they have been fitted alone. The fits were performed once with a one 
temperature collisionally-ionized diffuse gas and  
once with a cooling flow model (\textsc{APEC} and \textsc{mkcflow} in \textsc{Xspec}, respectively). We here
note that a similar result within uncertainties 
could be obtained using a \textsc{cemekl} model.
Then, we fitted the $ASCA$ spectra with the \textsc{APEC} model and \textsc{mkcflow}, with a complex absorber as necessary to achieve a reasonable fit. 
Then we re-fitted the $ASCA$ spectra fixing the temperature to the best-fit values first found for the BAT spectra alone. To be conservative, we used the range of luminosity obtained from the minimum and maximum fluxes among
the four spectral fits to the $ASCA$ spectra.  For the three sources simultaneously observed with \XMM\ and $NuSTAR$, the broad band spectrum was fitted with similar models. 
For J0927,  J2113 and \IGRJ\ we used instead their best 
fitting broad-band spectral models presented in \cite{bernardini17} and in this  work, respectively. 
We here note that for those IPs showing a soft X-ray blackbody component 
\citep[see][]{bernardini17}, the range of minimum and maximum bolometric fluxes encompass
this component.

Then luminosities were computed using distances reported in the literature. When available, the estimate of the binary inclination were also collected (see Table \ref{tab:orb_mod}).

\begin{table}
\caption{Main parameters of the IP sample: the orbital modulation depth in 
the 0.2--7 keV range, the bolometric (0.01--100 keV) luminosity (L), the distance (d), and binary inclination, when available. The instrument(s) (ins.) used to perform the observation is also reported, where A stands for $ASCA$, X+N for \XMM+$NuSTAR$, and X for \XMM.}
{\small
\begin{center} 
\tabcolsep=0.1cm
\begin{tabular}{llllll}
\hline  
Source       &  ins.     & depth   & L                & d    & $i^b$        \\
             &        &  \%    & 10$^{32}$ erg\,s$^{-1}$  & pc  & deg  \\         
\hline 
V1025 Cen    &  A      &  $3\pm3$       &   $2.12\pm^{2.2}_{1.4}$      & $230\pm^{70}_{70}$  [1]   & -        \\
\hline                                          
BG\,CMi       &  A      &  $82\pm9$      &   $42.0\pm^{17.9}_{13.3}$    & $553\pm^{75}_{86}$  [1]   &  55-75   \\
\hline                                          
V1223 Sgr    &  A      &  $28\pm3$     &   $127.5\pm^{56.6}_{37.8}$   & $527\pm^{54}_{43}$ [2]    & 16-40      \\
             &  X+N    &  $11\pm2$      &   $178.3\pm^{99.5}_{70.4}$   &                        &  \\
\hline                                          
V2400 Oph    &  A      &  $4\pm2$       &   $26.8\pm^{42.0}_{16.3}$    & $280\pm^{150}_{100}$ [2]  & 10:   \\
\hline                                          
AO Psc       &  A      & $47\pm5$       &   $20.1\pm^{35.9}_{13.0}$    & $330\pm^{180}_{120}$ [2]  & 60:    \\
\hline                                          
YY Dra       &  A      &  $9\pm4$       &   $1.1\pm^{0.7}_{0.6}$       & $155\pm^{35}_{35}$ [3]    & 42$\pm$5  \\
\hline                                          
V405 Aur     &  A      &  $6\pm2$       &   $26.0\pm^{59.8}_{17.5}$    & $380\pm^{210}_{130}$ [2]  &  -   \\
             &  A      &  $21\pm4$      &   $17.4\pm^{33.0}_{11.5}$    &                        &  \\
\hline                                          
FO Aqr       &  A      &  $102\pm13$    &   $61.5\pm^{106.0}_{39.7}$   & $450\pm^{240}_{160}$ [2]  & 65:  \\
\hline                                          
PQ Gem       &  A      &  $10\pm3$      &   $55.0\pm^{125.7}_{38.5}$   & $510\pm^{280}_{180}$ [2]  &  -   \\
             &  A      &   $3\pm4$      &   $34.6\pm^{67.6}_{23.7}$    &                        &   \\
\hline                                          
TV Col       &  A      &   $47\pm6$     &   $42.0\pm^{7.5}_{7.3}$      & $368\pm^{17}_{15}$ [2]    & 70:    \\
\hline                                          
TX Col       &  A      &   $58\pm6$     &   $43.0\pm^{59.1}_{28.7}$    & $591\pm^{135}_{175}$ [1]  & $<$25 \\
\hline                                          
V1062 Tau    &  A      &   $35\pm12$    &   $325.9\pm^{423.2}_{198.0}$ & $1400\pm^{700}_{500}$[2]  & -       \\
\hline                                          
V709 Cas     &  X+N    &   $19\pm1$     &   $7.0\pm^{1.6}_{1.4}$       & $230\pm^{20}_{20}$   [4]  & -  \\
\hline 
NY Lup       &  X+N    &   $<10$        &   $100.8\pm^{108.5}_{54.8}$  & $690\pm^{150}_{150}$ [3]  & 25-58    \\
\hline 
J0927        &  X      & $91\pm2$       &   $14.5\pm^{20.2}_{10.1}$    & $670\pm^{268}_{268}$ [5]$^a$  & - \\
\hline 
J14257       &  X      & $95\pm3$       &   $8.3\pm^{16.2}_{6.3}$      & $563\pm^{225}_{225}$ [6]$^a$  & - \\
\hline 
J2113        &  X      & $76\pm1$       &   $19.5\pm^{29.3}_{14.4}$    & $750\pm^{300}_{300}$ [5]$^a$  & - \\   
\end{tabular}
\label{tab:orb_mod} 
\end{center}}
\begin{flushleft}
$^a$ We assumed a 40 per cent uncertainty on the distance. \\
$^b$ See Koji Mukai Website and more reference therein (https://asd.gsfc.nasa.gov/Koji.Mukai/iphome/iphome.html).\\
\noindent Distances from: [1] \cite{ak08} [2]; \cite{parker05}; [3] 
Koji Mukai Website; 
[4] \cite{Bonnet-Bidaud01}; [5] \cite{bernardini17}; [6] This work. Uncertain distances are reported with a side colon.\\
\end{flushleft}  
\end{table}

\section*{Acknowledgments}

FB is founded by the European Union's Horizon 2020 research and innovation 
programme under the Marie Sklodowska-Curie grant agreement n. 664931. DdM 
acknowledges financial support from the Italian Space Agency and National 
Institute for Astrophysics, ASI/INAF, under agreements 
ASI-INAF I/037/12/0 and and ASI-INAF n.2017-14-H.0.
This work is based on observations obtained with \XMM , an ESA science mission 
with instruments and contributions directly funded by ESA Member States; 
with \Swift, a National Aeronautics and Space Administration (NASA) science 
mission with Italian participation. This work has also made use of the
the Two Micron All Sky Survey (2MASS), a joint project of the University
of Massachusetts and the Infrared Processing and Analysis
Center (IPAC)/Caltech, funded by NASA and the NSF. 

\bibliographystyle{mn2e}
\bibliography{biblio}

\vfill\eject
\end{document}